\documentclass[lettersize,journal]{IEEEtran}
\usepackage{amsmath,amsfonts}
\usepackage{algorithmic}
\usepackage{algorithm}
\usepackage{array}
\usepackage[caption=false,font=normalsize,labelfont=sf,textfont=sf]{subfig}
\usepackage{textcomp}
\usepackage{url}
\usepackage{verbatim}
\usepackage{graphicx}
\usepackage{cite}
\usepackage{float}
\usepackage{stfloats}
\usepackage{cuted}
\usepackage{caption}
\usepackage{lettrine}
\usepackage{cite}
\begin{document}

\title{Adaptive Wavelet Division Multiplexing for Multiple Users with Heterogeneous Mobility Profiles over Rayleigh Fading Channels}

\author{%
{Seyyed Erfan Fatemi},~
{Wafa Hedhly},~\textit{Member, IEEE},~
{Leila Musavian},~\textit{Senior Member, IEEE},~and~
{Nikolaos Thomos},~\textit{Senior Member, IEEE}%
\thanks{The authors are with the School of Computer Science and Electronic Engineering, University of Essex, Wivenhoe Park, Colchester CO4 3SQ, United
Kingdom (e-mail: {s.fatemi, wafa.hedhly, leila.musavian, nthomos}@essex.ac.uk). This work was funded by the Engineering and Physical Science Research Council, UK, [grants number EP/X012204/1, EP/X04047X/2, EP/Y037243/1].}}

\markboth{IEEE Communications Letters}%
{Fatemi \MakeLowercase{\textit{et al.}}: Adaptive Wavelet Division Multiplexing for Heterogeneous Mobility Users}

\maketitle

\begin{abstract}
This paper proposes an adaptive wavelet division multiplexing scheme for wireless systems serving users with heterogeneous mobility profiles over frequency-selective Rayleigh fading channels. By exploiting the multiresolution structure of the discrete wavelet transform (DWT), users are adaptively assigned to different decomposition levels according to their channel dynamics and Doppler conditions. A single-tap minimum mean square error (MMSE) equalizer is applied in the frequency domain, and the system performance is evaluated under realistic time-varying multipath fading environments. Simulation results demonstrate that the proposed adaptive allocation achieves balanced bit error rate (BER) across all user mobility classes while delivering substantial peak-to-average power ratio (PAPR) reductions relative to both conventional orthogonal frequency division multiplexing (OFDM) and orthogonal time-frequency space (OTFS) modulation. The proposed framework is further validated in a four-user heterogeneous-mobility scenario, confirming its scalability and effectiveness to mixed-mobility multi-user scenarios.
\end{abstract}


%----------------------------------------------------------------------
\section{Introduction}
%----------------------------------------------------------------------

\lettrine{O}RTHOGONAL frequency division multiplexing (OFDM) remains the dominant multicarrier waveform in contemporary wireless standards due to its inherent robustness to frequency-selective multipath channels. However, its reliance on Fourier basis functions of infinite temporal support leads to poor time localization, high peak-to-average power ratio (PAPR), and severe sensitivity to Doppler-induced inter-carrier interference (ICI) under time-varying channels~\cite{oltean2007wavelet,oltean2010wavelet}.

Wavelet-based multicarrier modulation, commonly referred to as Wavelet-OFDM (W-OFDM), has emerged as a promising alternative built upon multiresolution analysis and orthogonal filter-bank theory~\cite{mallat2002theory,vetterli2002wavelets}. A substantial body of literature has shown that W-OFDM can match or surpass OFDM under both flat and frequency-selective Rayleigh fading and remains effective in practical deployments~\cite{oltean2007wavelet,oltean2010wavelet,chafii2017wavelet,10670341}. Moreover, W-OFDM exhibits a scale-dependent BER behavior arising from the distinct time-frequency tiling associated with each wavelet scale, and admits substantially lower PAPR statistics~\cite{learned1994wavelet,kucur1998wavelet,khan2018wavelet}. These PAPR gains become more pronounced when employing short-support wavelets and lower number of decomposition levels~\cite{chafii2017wavelet}. In addition, multiuser multiple-input multiple-output OFDM (MIMO-OFDM) studies have shown that replacing the Fourier core with a wavelet filter bank improves error-rate performance while remaining compatible with low-complexity equalization techniques~\cite{hasan2012performance,ramadan2025enhanced}.

For high-mobility scenarios, alternative waveforms have been developed to mitigate Doppler dispersion and rapid time variations that severely degrade conventional OFDM through ICI. Among these, orthogonal time-frequency-space (OTFS) modulation exploits the full diversity of doubly selective channels, but it generally requires computationally intensive two-dimensional detection algorithms such as message-passing detectors or rake decision-feedback equalizers based on maximal-ratio combining. The complexity of these receivers scale with the number of resolvable channel paths and substantially exceeds that of single-tap OFDM equalization~\cite{9293173}. Multi-user OTFS extensions, such as localized orthogonal multiple access (LOMA) based on cyclic-prefix OTFS~\cite{10543767}, improve user separation in the time-frequency plane while still inheriting the higher complexity of OTFS detection. Affine frequency division multiplexing (AFDM), which multiplexes data onto chirp subcarriers via the discrete affine Fourier transform, preserves symbol orthogonality in the delay-Doppler domain and is well-suited to high-mobility channels~\cite{sabuj2025complex}. Recent work has further shown that replacing the Fourier core of AFDM with a complex discrete wavelet transform improves BER by over 35\% relative to conventional AFDM in LEO satellite scenarios, while simultaneously reducing transceiver complexity~\cite{sabuj2025complex}.

Unlike fixed-resolution transforms such as the FFT, the DWT enables adaptive multiresolution decompositions in the time-frequency domain, which is particularly beneficial in heterogeneous-mobility environments where high-mobility users require stronger time localization to track rapid channel variations, whereas low-mobility users benefit from enhanced frequency resolution and improved spectral compactness. However, existing wavelet-based multiple-access approaches~\cite{arshad2025wavelet,arshad2025deep} do not dynamically match the transform resolution to each user's Doppler profile, which is the key gap addressed in this paper.

Motivated by these limitations, this paper proposes an adaptive wavelet-based multiplexing scheme, in which users are assigned to different DWT decomposition levels according to their mobility profile. Specifically, we introduce a general $N$-user framework, 
%that allocates each user a decomposition level commensurate with its Doppler profile, 
where high-mobility users employ lower decomposition levels for superior time localization and reduced PAPR, whereas low-mobility users exploit higher decomposition levels for enhanced frequency resolution and spectral compactness. Multiple users may share the same decomposition level through the allocation of disjoint subsets of approximation and detail wavelet coefficients. Because of the joint time-frequency localization of the wavelet basis, the corresponding user signals occupy distinct regions of the time-frequency plane, thereby mitigating inter-user interference. We further provide a comprehensive BER and PAPR analysis across multiple wavelet families and decomposition levels under realistic mixed-mobility Rayleigh fading. The proposed scheme is compared against conventional OFDM and against zero-padded OTFS employing the same single-tap MMSE detector~\cite{9293173} ensuring that the comparison isolates waveform-dependent performance gains from receiver-specific processing advantages.

%----------------------------------------------------------------------
\section{Wavelet Modulation}
%----------------------------------------------------------------------
 \begin{figure*}[t]
\centering
 
\begin{minipage}[t]{0.48\textwidth}
    \centering
    \includegraphics[width=1.2\linewidth]{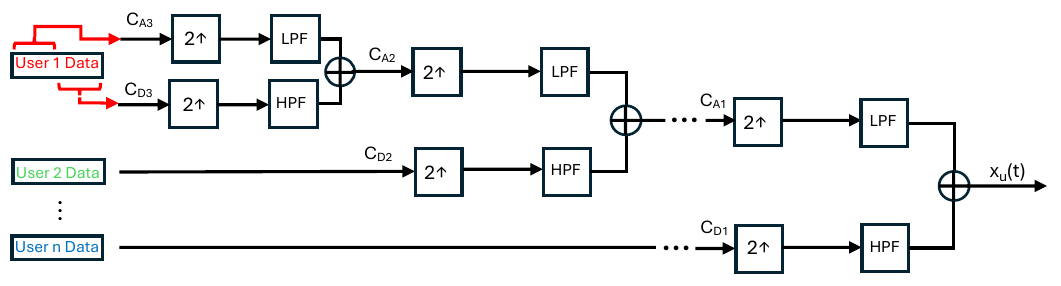}
    
    (a) n-Level IDWT Decomposition
\end{minipage}
\hfill
\begin{minipage}[t]{0.48\textwidth}
    \centering
    \includegraphics[width=0.6\linewidth]{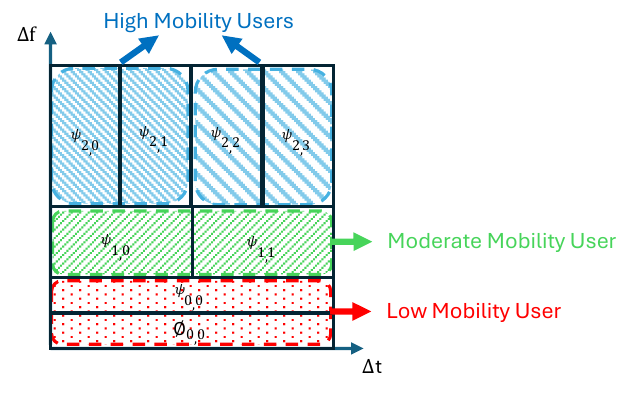}
    
    (b) Accommodating 4 Users in Time-frequency Grid 
\end{minipage}
 
\caption{Adaptive user placement in the time--frequency grid.}
\label{fig:combined}
\end{figure*}

Wavelets provide an orthogonal set of localized basis functions capable of representing signals jointly in time and frequency. Let $\psi(t)$ and $\phi(t)$ denote the mother wavelet and scaling functions, respectively, where $\psi(t), \psi(t)$ are in $L^2(\mathbb{R})$, the space of square-integrable functions. Their scaled and translated versions at decomposition level $j$ and position $k$ are given by
\begin{align}
\psi_{j,k}(t) &= 2^{j/2}\psi(2^{j} t - kT_0), \\
\phi_{j,k}(t) &= 2^{j/2}\phi(2^{j} t - kT_0),
\end{align}
where $T_0$ denotes the symbol duration. For compactly supported wavelets, $\psi_{j,k}$ and $\phi_{j,k}$ have support of $\bigl[kT_0/2^{j},\,(k{+}1)T_0/2^{j}\bigr]$. Hence, increasing the decomposition level $j$ reduces the temporal support by a factor of two, while correspondingly doubling the spectral occupancy ~\cite{chafii2017wavelet}. Wavelet and scaling functions $\{\psi_{j,k},\phi_{j,k}\}$ together form an orthonormal basis of $L^2(\mathbb{R})$ over the selected range of scales~\cite{mallat2002theory}. 

The choice of mother wavelet determines the analytical shape of $\psi(t)$ and $\phi(t)$ and hence associated time-frequency localization properties. The decomposition level $L_g$ controls the time-frequency tiling: a small $L_g$ preserves the finest time localization, whereas a larger $L_g$ further splits the low-frequency band into narrower subbands, improving frequency resolution at the cost of time resolution.

The DWT and the inverse DWT (IDWT)  are realized through a two-channel quadrature mirror filter bank with synthesis low-pass and high-pass filters $\{h_l,h_h\}$ and their corresponding analysis filters $\{g_l,g_h\}$ of length $K$, all derived from the selected wavelet~\cite{mallat2002theory,khan2018wavelet}. By iteratively applying these filters across $L_g$ decomposition levels, together with dyadic downsampling in the analysis stage and dyadic upsampling in the synthesis stage, the complete multiresolution transform is obtained.

%----------------------------------------------------------------------
\section{System Model}
%----------------------------------------------------------------------
\begin{figure*}[t]
  \centering
  \includegraphics[width=\textwidth]{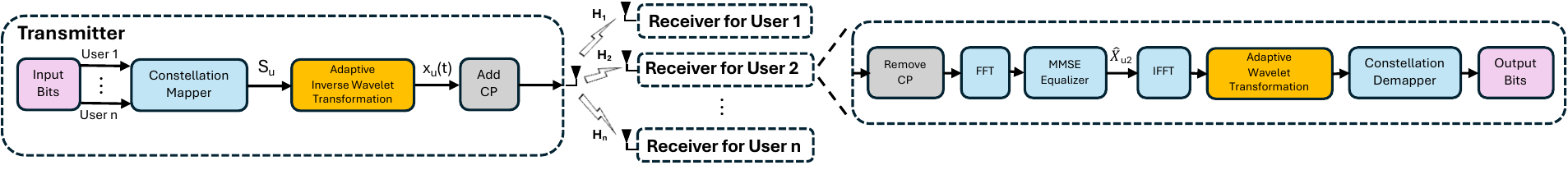}
  \caption{System model of the proposed scheme with $N$ users.}
  \label{fig:wide_userplacement}
\end{figure*}

We consider a multi-user communication system serving $U$ users over independent frequency-selective Rayleigh fading channels, where each user is characterized by a distinct Doppler profile. The users are grouped into $G$ mobility classes
$\mathcal{G}_1,\mathcal{G}_2,\ldots,\mathcal{G}_G$, ordered from lowest to highest mobility. Each class $g$ is assigned a decomposition level $L_g$ such that $L_1 \geq L_2 \geq \cdots \geq L_G$, reflecting the principle that low-mobility users tolerate higher decomposition levels, whereas higher-mobility users operate at lower decomposition levels to preserve time resolution and enhanced Doppler resilience.

The binary message of user $u\in\mathcal{G}_g$ is first mapped onto complex-valued quadrature amplitude modulation (QAM) symbols $\mathbf{S}_u\in\mathbb{C}^{N_u}$, where $N_u$ denotes the number of data symbols allocated to user $u$. Subsequently, these symbols are assigned to the user-specific subsets of approximation coefficients $a^{(u)}_{J-L_u,q}[n]$ on the coarsest scale $J-L_u$ and detail coefficients $w^{(u)}_{j,k}[n]$ on scale $j$ and position $k$. The continuous-time signal of the user $u$ is then synthesized through an $L_u$-level IDWT realized through a two-channel filter-bank structure structure as illustrated in Fig.~\ref{fig:combined}(a) and  expressed as

\begin{equation}
\begin{split}
x_u(t) = &\sum_{n}\sum_{j=J-L_u}^{J-1}\sum_{k=0}^{2^j-1}
        w^{(u)}_{j,k}[n]\,\psi_{j,k}(t-nT_0) \\
       &+ \sum_{n}\sum_{q=0}^{2^{J-L_u}-1}
         a^{(u)}_{J-L_u,q}[n]\,\phi_{J-L_u,q}(t-nT_0),
\end{split}
\end{equation}
where n denotes the discrete wavelet symbol index, and $\psi_{j,k}(t-nT_0)$ and $\phi_{J-L_u,q}(t-nT_0)$ denote wavelet and scaling functions, respectively. Multiple users may share the same decomposition level, with each user assigned to a disjoint subset of approximation and detail coefficient indices. Due to the time-frequency localization of the wavelet basis, the corresponding signals occupy separate localized regions in the time-frequency plane, limiting this way inter-user interference. This adaptive allocation enables a mobility-aware tradeoff between temporal and frequency localization, tailored to the channel dynamics experienced by each user, as illustrated in Fig.~\ref{fig:combined}(b).

Following wavelet-based signal synthesis, a cyclic prefix (CP) of length $N_\mathrm{cp}$ is
prepended to each transmission frame in order to mitigate inter-symbol interference (ISI) caused by
multipath propagation. Although a W-OFDM implementation could omit the
CP \cite{hasan2012performance}, we retain it here to enable a low-complexity single-tap equalization at the receiver. The composite transmitted signal is expressed as follows
\begin{equation}
x(t) = \sum_{u=1}^{N} x_u(t).
\end{equation}

The propagation channel is modeled as a linear time-varying frequency-selective Rayleigh fading channel comprising $P$ propagation paths. Accordingly, the baseband channel impulse response experienced by user $u$ is expressed as 
\begin{equation}
h_u(t,\tau) = \sum_{p=1}^{P} \beta_{u,p}\,e^{j 2\pi \nu_{u,p} t}\,
              \delta(\tau - \tau_p),
\end{equation}
where $\beta_{u,p}\sim\mathcal{CN}(0,\sigma_p^2)$ is the complex Gaussian gain of the $p$-th path, $\tau_p$ stands for its delay, and $\nu_{u,p} = f_{d_u}\cos(\theta_{u,p})$ is the Doppler shift induced by the relative motion of the user $u$, with $f_{d_u}$ denoting the maximum Doppler frequency and $\theta_{u,p}$ the arrival angle of the $p$-th path. The discrete-time signal observed by user $u$, after sampling and multipath convolution of the composite waveform is given by
\begin{equation}
y_u[n] = \sum_{p=1}^{P} \beta_{u,p}\,e^{j 2\pi \nu_{u,p} n T_s}\,
         x[n-\ell_p] + \eta[n],
\end{equation}
where $T_s$ corresponds to the sample duration, $\ell_p$ is the $p$-th path delay in samples, and $\eta[n]\sim\mathcal{CN}(0,N_0)$ is the additive white Gaussian noise.

At the receiver, as illustrated in Fig.~\ref{fig:wide_userplacement}, the CP is first removed and then channel equalization is performed in the frequency domain. This approach exploits the fact that the synthesized wavelet sequence is processed by the same circulant CP-protected channel as a conventional OFDM signal \cite{chafii2017wavelet,hasan2012performance}. Concretely, an $N$-point FFT is applied to the CP-stripped received block resulting in
\begin{equation}
Y_u[k] = H_u[k]\,X_u[k] + \eta_u[k],
\end{equation}
where $X_u[k]$ and $Y_u[k]$ are the FFTs of the CP-stripped transmitted and received blocks of user $u$, $H_u[k]$ is the user's channel frequency response at subcarrier $k$, and $\eta_u[k]$ is the corresponding frequency-domain noise sample. A single-tap MMSE equalizer is applied to each subcarrier as shown below
\begin{equation}
\hat{X}_u[k] = \frac{H_u^*[k]}{|H_u[k]|^2 + 1/\gamma}\,Y_u[k],
\end{equation}
where $\gamma$ denotes the transmit signal-to-noise ratio (SNR) per subcarrier. The equalized frequency-domain block is transformed back to the time domain via an inverse FFT(IFFT), producing the recovered samples
$\hat{x}^{(u)}[n]=\mathrm{IFFT}\{\hat{X}_u[k]\}$. The original QAM symbols are then extracted by recursively applying the analysis filter bank corresponding to the user's $L_u$-level DWT \cite{mallat2002theory,khan2018wavelet}
\begin{equation}
\begin{split}
\hat{a}^{(u)}_{j,k}[n] &= \sum_{m} g_l[m-2k]\,\hat{x}^{(u)}[n], \\
\hat{w}^{(u)}_{j,k}[n] &= \sum_{m} g_h[m-2k]\,\hat{x}^{(u)}[n],
\end{split}
\end{equation}
where
\begin{equation}
g_l[n] = h_l[-n], \qquad g_h[n] = h_h[-n],
\end{equation}
and
\begin{equation}
g_h[n] = (-1)^{n}\,g_l[L_f-1-n],
\end{equation}
where $L_f$ stands for the filter length set by the mother wavelet. The estimated symbols $\hat{\mathbf{S}}_u$ are then read out from the corresponding subset of $\{\hat{a}^{(u)}_{j,k}[n],\,\hat{w}^{(u)}_{j,k}[n]\}$ allocated to user $u$; orthogonality of the wavelet basis ensures that contributions from the other users in the composite signal $x$ are mapped onto disjoint coefficient indices and therefore do not interfere with the recovery process.

%----------------------------------------------------------------------
\section{Simulation Results}
%----------------------------------------------------------------------
We evaluate the performance of the proposed W-OFDM based multiplexing scheme in terms of bit error rate (BER) and the complementary cumulative distribution function (CCDF) of the PAPR of the transmitted waveform. The propagation environment is modeled using the Extended Typical Urban (ETU) channel profile of Table~\ref{tab:etu}, and the remaining system parameters are listed in Table~\ref{tab:params}.

\begin{table}[t]
\centering
\caption{ETU Channel Delay and Power Profile}
\label{tab:etu}
\resizebox{\columnwidth}{!}{%
\begin{tabular}{lcccccccccc}
\hline
\textbf{Delay (ns)}         & 0   & 50   & 120  & 200 & 230 & 500 & 1600 & 2300 & 5000 \\
\textbf{Avg.\ Gain (dB)}    & $-$1 & $-$1 & $-$1 & 0   & 0   & 0   & $-$3 & $-$5 & $-$7 \\
\hline
\end{tabular}}
\end{table}

\begin{table}[t]
\centering
\caption{Simulation Parameters}
\label{tab:params}
\begin{tabular}{|l|c|}
\hline
\textbf{Parameter} & \textbf{Value} \\
\hline
FFT / IDWT size & 128 \\
Cyclic Prefix Length & 32 \\
Modulation & 4-QAM \\
Subcarrier Spacing & 15~kHz \\
SNR Range & 0--25~dB \\
Maximum Doppler Shifts & 300~Hz \\
Monte Carlo Trials & $10^6$ \\
Wavelet Family (default) & Daubechies-4 (db4) \\
\hline
\end{tabular}
\end{table}

\begin{figure}[t]
\centering
\includegraphics[width=\columnwidth]{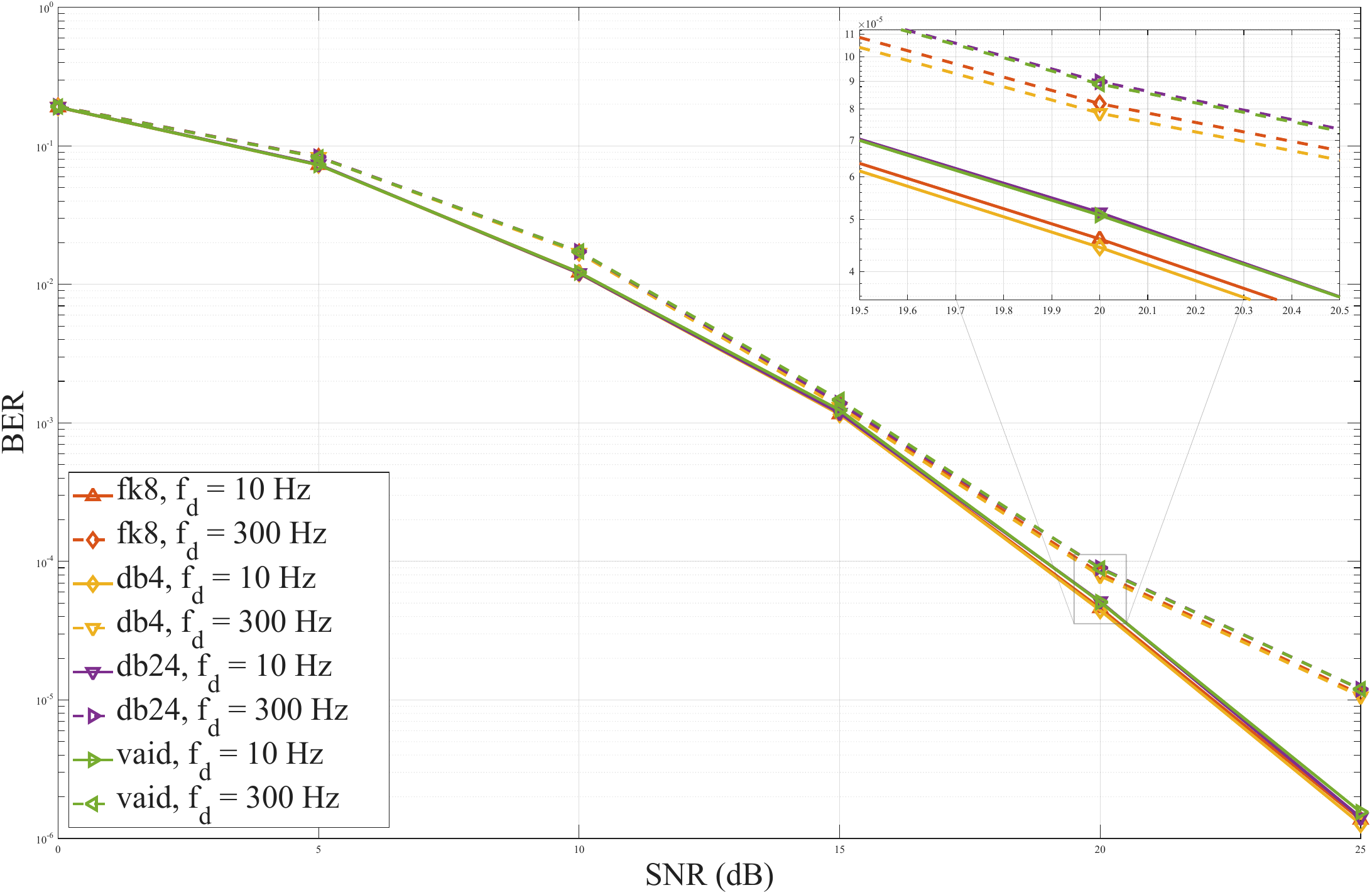}
\caption{BER performance comparison of various wavelet families.}
\label{fig:userplacement}

\vspace{0.5em}

\includegraphics[width=\columnwidth]{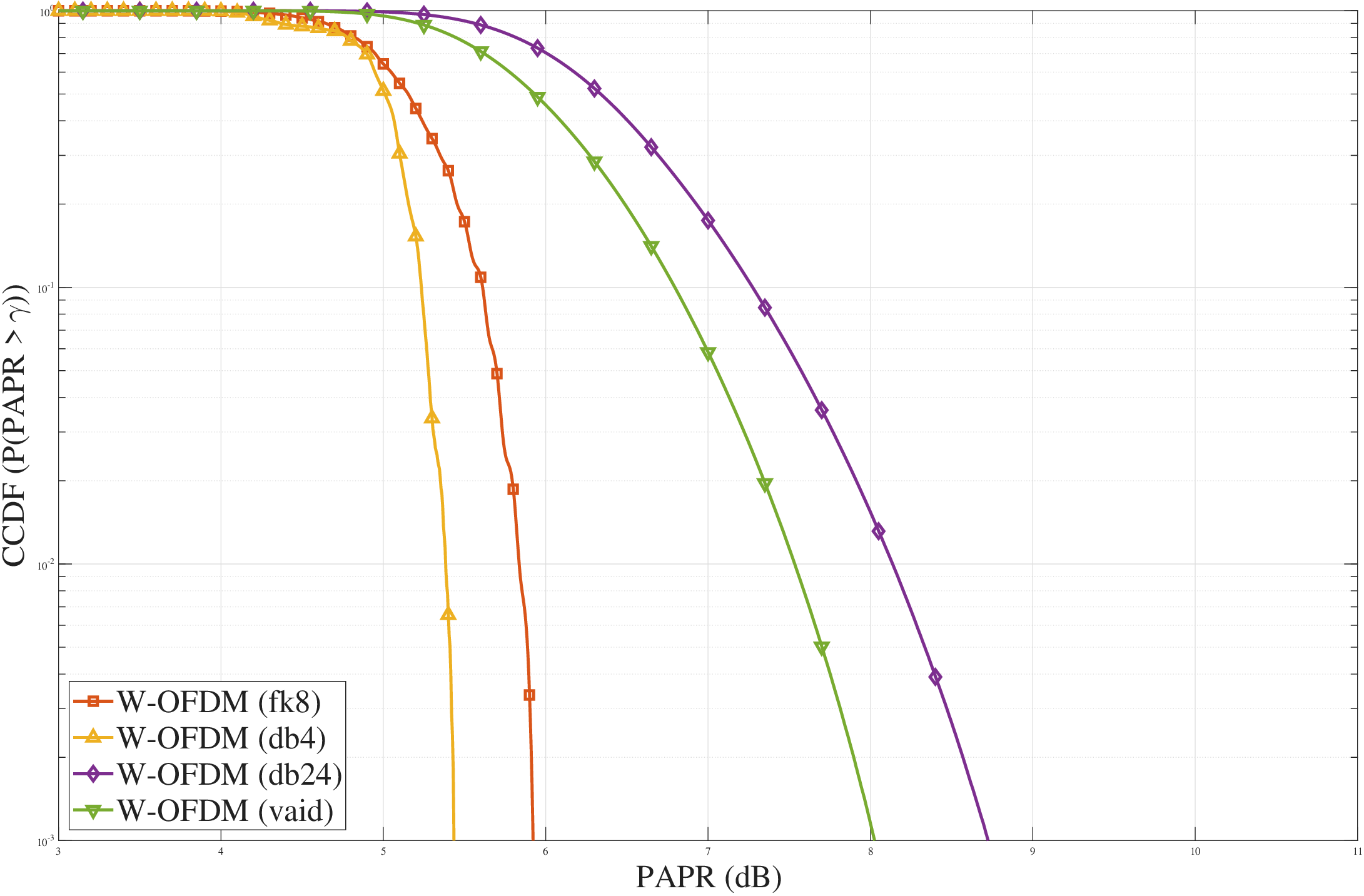}
\caption{PAPR performance comparison of various wavelet families.}
\label{fig:Fatemi_PAPR_WF}
\end{figure}

We start first by evaluating a single-user configuration using four representative wavelet families: Fej\'{e}r-Korovkin (fk8), Daubechies (db4 and db24), and Vaidyanathan (Vaid). The BER and PARP performance are evaluated for various SNR values and the results are shown in Fig.~\ref{fig:userplacement} and Fig.~\ref{fig:Fatemi_PAPR_WF}, respectively. Although the BER curves are nearly indistinguishable across all wavelet families, the PAPR characteristics differ noticeably. This observation is consistent with~\cite{chafii2017wavelet} which reports that shorter-support wavelets such as db4 and fk8 achieve the lowest PAPR because of their superior time localization and reduced amplitude fluctuations. Based on the overall trade-off between BER and PAPR, db4 is employed by our scheme for all subsequent simulations.

\begin{figure}[t]
\centering
\includegraphics[width=\columnwidth]{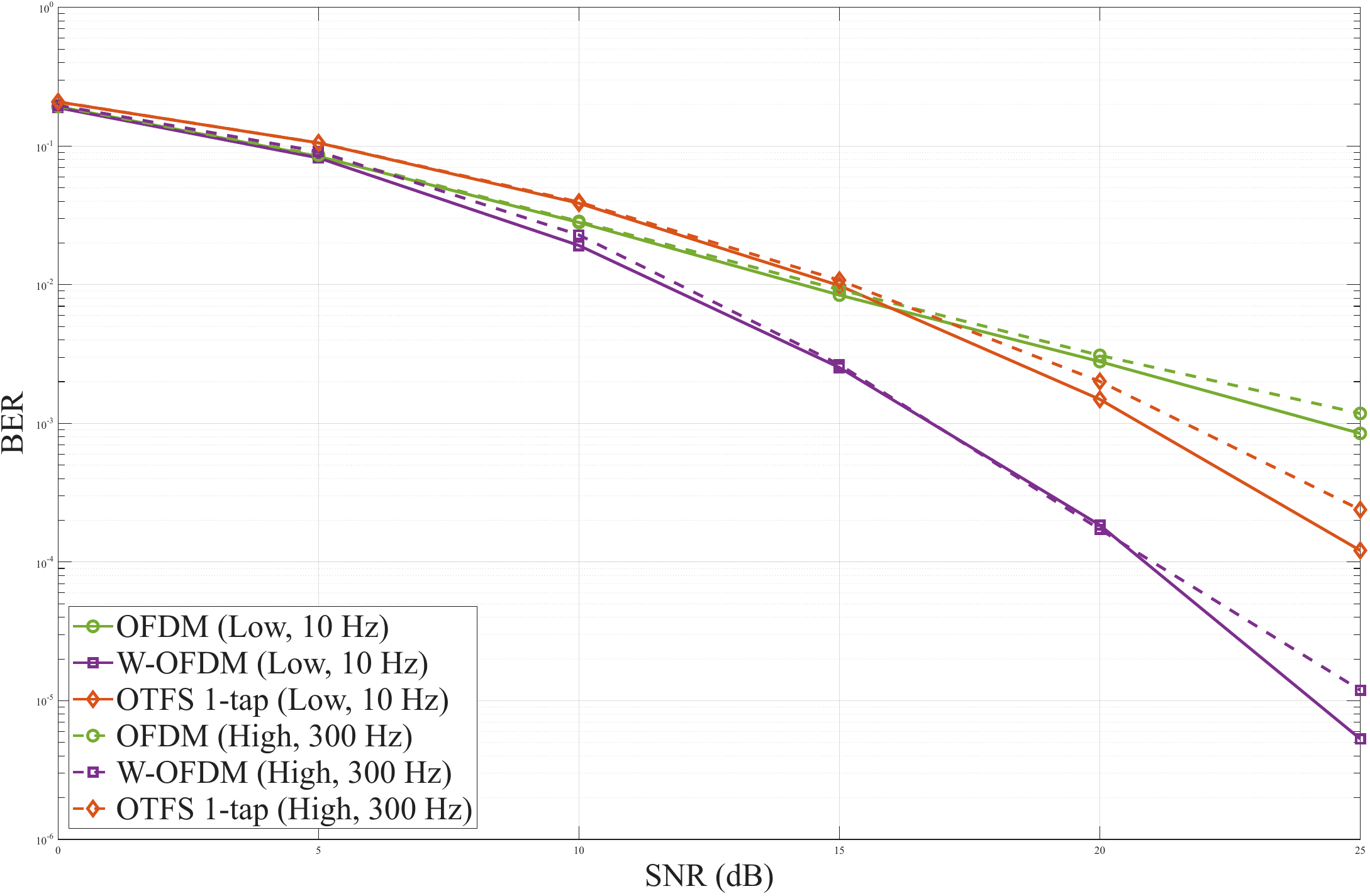}
\caption{BER performance comparison of OFDM, OTFS and W-OFDM.}
\label{fig:Fatemi_BER_all3}

\vspace{0.5em}

\includegraphics[width=\columnwidth]{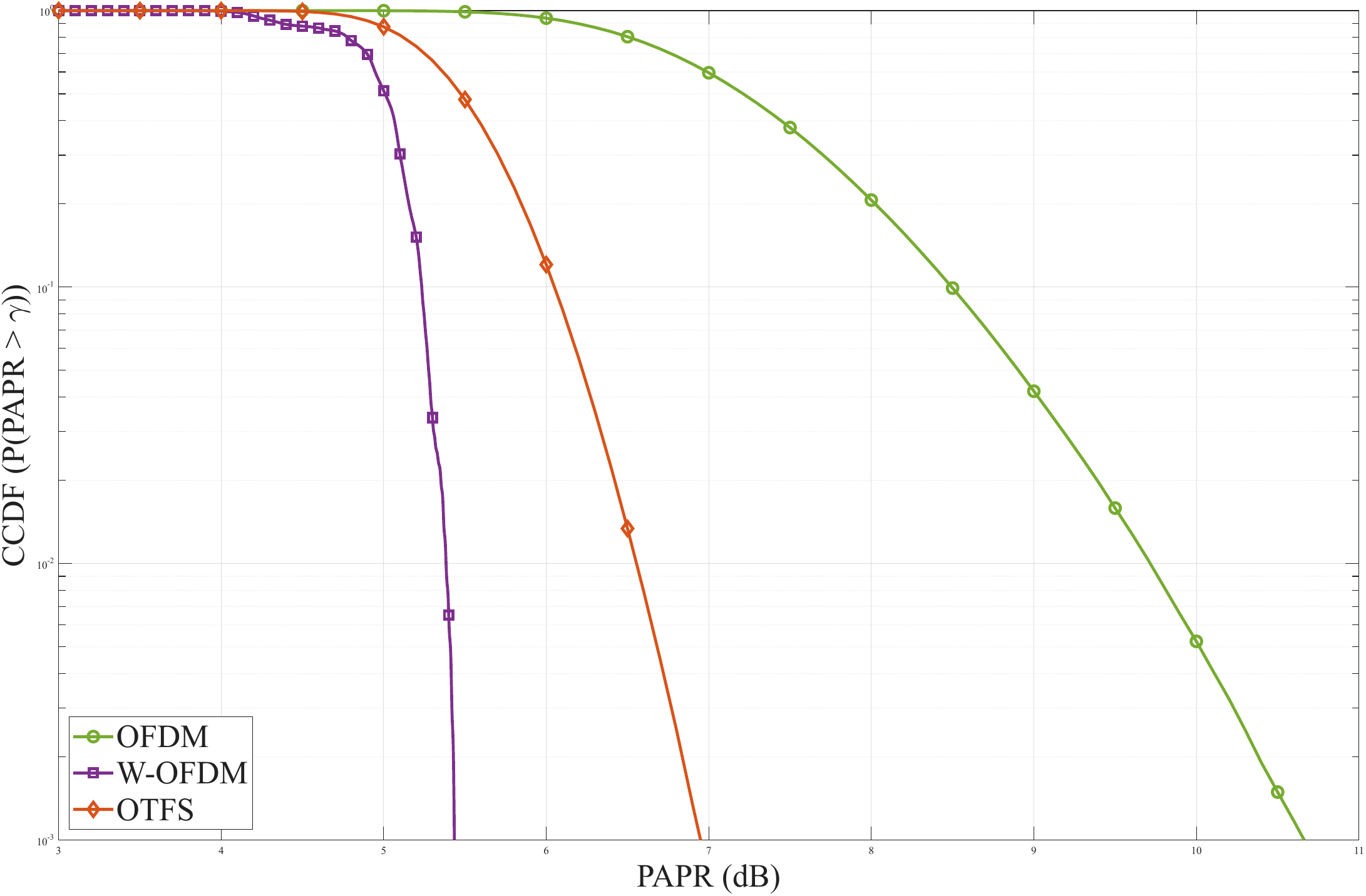}
\caption{PAPR performance comparison of OFDM, OTFS and W-OFDM.}
\label{fig:Fatemi_PAPR_all3}
\end{figure}

Next, we examine the BER performance of the proposed scheme over range of various SNR values and compare it against OFDM and OTFS in a two-user heterogeneous mobility scenario comprising a low-mobility user (pedestrian) and a high-mobility user (vehicle). The corresponding results are depicted in Fig.~\ref{fig:Fatemi_BER_all3}. To ensure a fair complexity comparison, all three waveforms are equalized with the same single-tap frequency-domain MMSE detector, while the OTFS implementation follows the rectangular-pulse zero-padded framework of~\cite{9293173}. From this comparison, we can observe that the proposed scheme outperforms both OFDM and OTFS for both mobility profiles. This performance gain arises because OFDM suffers significantly from ICI, whereas OTFS, whose information symbols span the entire time-frequency plane, cannot deliver its full diversity gain without the iterative two-dimensional detectors proposed in~\cite{9293173}. The corresponding PARP performance for all considered waveforms is shown in Fig.~\ref{fig:Fatemi_PAPR_all3}. As observed from this figure, W-OFDM achieves substantially lower PARP than both OFDM and OTFS because of its superior time localization properties and the localized allocation of users across wavelet subbands.

To assess the effectiveness of the proposed mobility-aware decomposition-level assignment, we further consider a four-user scenario with distinct Doppler profiles: $f_{d_1}=10$~Hz (low mobility), $f_{d_2}=100$~Hz (moderate mobility), $f_{d_3}=200$~Hz (high mobility), and $f_{d_4}=300$~Hz (high mobility). Two allocations, denoted as Scenario-1 and Scenario-2, are investigated in order to examine the impact of user-to-decomposition-level assignment on system performance. In Scenario~1, the low-mobility user is assigned to the third decomposition level, the moderate-mobility user to the second decomposition level, and the two high-mobility users are assigned to the first decomposition level, i.e., $\{L_1,L_2,L_3,L_4\}=\{3,2,1,1\}$. In contrast Scenario~2 adopts the reverse assignment strategy, i.e., $\{L_1,L_2,L_3,L_4\}=\{1,1,2,3\}$. 

\begin{figure}[t]
\centering
\includegraphics[width=\columnwidth]{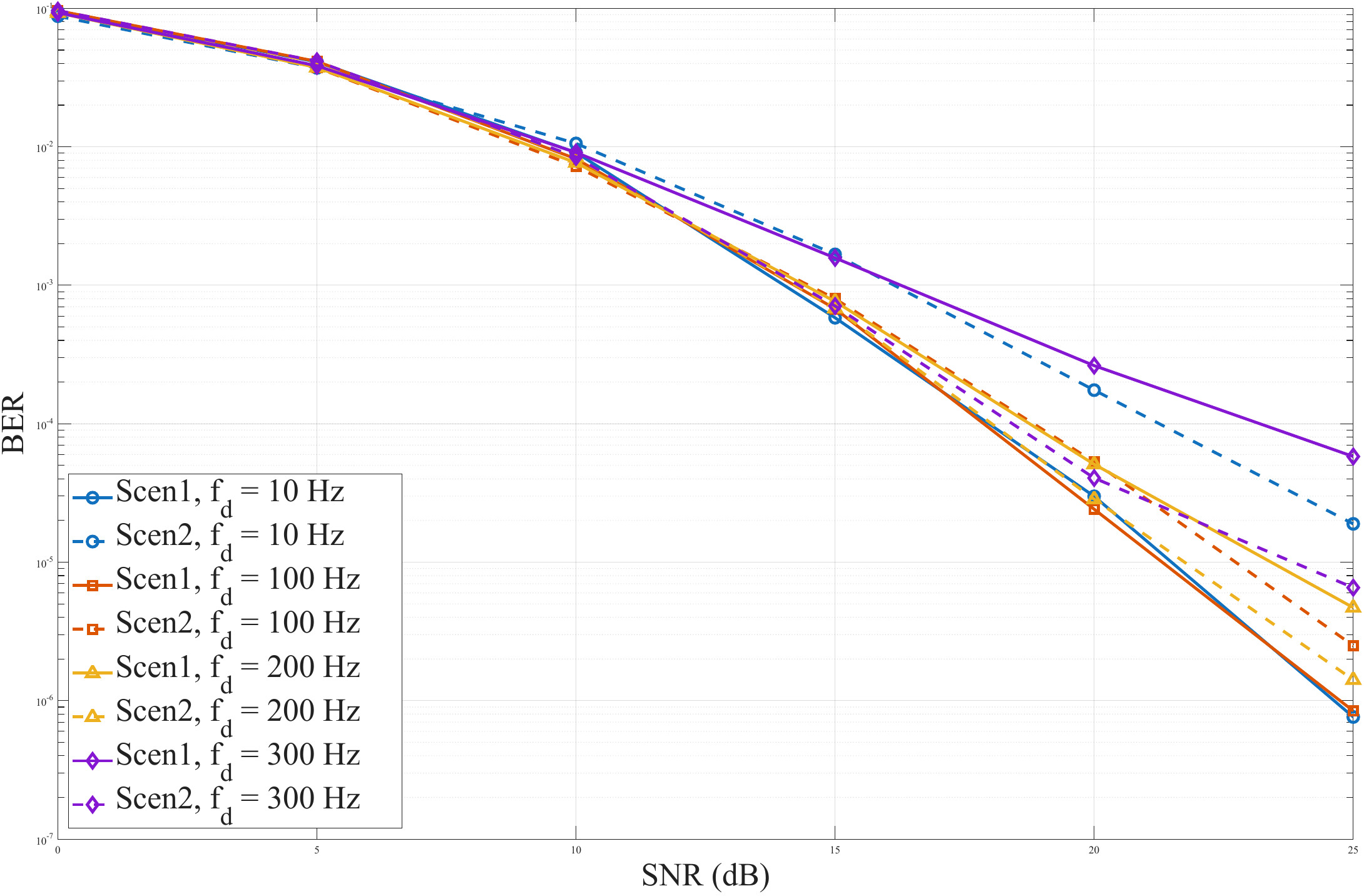}
\caption{BER performance comparison for two allocation scenarios in a four-user heterogeneous-mobility system.}
\label{fig:Fatemi_BER_2Scenarios}
\end{figure}

The BER performance comparison of the two allocation strategies for various SNR values is illustrated in Fig.~\ref{fig:Fatemi_BER_2Scenarios}. The observed BER behavior reflects how the multiresolution wavelet structure interacts with users experiencing different mobility conditions. At decomposition level $L_g$, each coefficient corresponds to a basis function of duration $2^{L_g}T_0$ and bandwidth $B/2^{L_g}$. Therefore, increasing $L_g$ reduces the bandwidth associated with each coefficient while increasing the time interval over which the corresponding basis function must remain coherent with the channel. For low-mobility users, the channel coherence time remains significantly larger than $2^{L_g}T_0$ even at $L_g=3$. As a result, the longer-duration basis functions stay coherent and the narrow subbands experience nearly flat fading,  improving error resilience without incurring additional bandwidth consumption. In contrast, for high-mobility users, the channel varies substantially within a large-$L_g$ symbol, so restricting them to $L_g=1$ preserving within symbol coherence and also providing wider per-coefficient bandwidths that are inherently more robust to ICI. Scenario~2 reverses this mobility-aware assignment strategy, leading to the BER degradation observed in this figure.

The clear separation between the two curves in Fig.~\ref{fig:Fatemi_BER_2Scenarios} empirically validates the central design principle of this letter: high-Doppler users systematically achieve better BER at low decomposition levels, whereas low-Doppler users benefit from higher decomposition levels. Hence, the proposed adaptive scheme provides a more balanced BER performance across heterogeneous mobility classes. This means that no user is forced to operate at a substantially higher error rate than the others under a common SNR. This property is particularly desirable in practical multi-user deployments whose link-level quality of service is typically constrained by the worst user. By improving the BER performance of high-mobility users to a similar error level as low-mobility users directly improves the system-level outage performance. It is worth noting that the proposed scheme also consistently outperforms OFDM, particularly at high Doppler where the Fourier-based waveforms suffer from severe ICI degradation.

%----------------------------------------------------------------------
\section{Conclusion}
%----------------------------------------------------------------------

This paper has proposed an adaptive wavelet division multiplexing scheme in which the DWT decomposition level is matched to the mobility regime of each user
%---one level for high-mobility, two levels for moderate-mobility, and three levels for low-mobility users---
achieving balanced BER across heterogeneous mobility classes while simultaneously offers substantial PAPR reductions relative to OFDM and OTFS under a common single-tap MMSE equalizer. The performance gains arise from the multiresolution structure of the wavelet basis, which adapts the time-frequency tiling to the channel coherence time of each user and provides a favourable middle ground between ICI-limited OFDM and complexity-limited OTFS~\cite{9293173}. Future work will explore advanced equalization and multi-antenna extensions.

%\printbibliography
\bibliographystyle{IEEEtran}
\bibliography{Main}

%\vspace{11pt}

\end{document}